\documentclass[preprint2]{aastex6}

\usepackage[normalem]{ulem}

\usepackage{graphicx}   
\usepackage{dcolumn}    
\usepackage{bm}         
\usepackage{verbatim}   
\usepackage{booktabs}
\usepackage{amsmath,amssymb}
\usepackage{color}
\usepackage{multirow}
\usepackage{inputenc}
\usepackage{natbib}
\bibpunct{(}{)}{;}{a}{}{,}
\usepackage{floatrow}
\usepackage[caption=false]{subfig}
\usepackage{savesym}
\savesymbol{tablenum}
\usepackage{siunitx}
\restoresymbol{SIX}{tablenum}
\usepackage{enumitem}   

\renewcommand\leq{\leqslant}
\renewcommand\geq{\geqslant}
\usepackage[normalem]{ulem}
\newcommand\bsout{\bgroup\markoverwith{\textcolor{blue}{\rule[0.5ex]{2pt}{0.4pt}}}\ULon}

\begin{document}
\title{Synergy of stochastic and systematic energization of plasmas
during
turbulent reconnection}
\author{Theophilos Pisokas, Loukas Vlahos and Heinz Isliker}
\affiliation{Department of Physics, Aristotle University of Thessaloniki\\
GR-52124 Thessaloniki, Greece}

\date{\today}
\begin{abstract}
The important characteristic of turbulent reconnection is that it combines large scale magnetic disturbances $(\delta B/B \sim 1)$ with randomly distributed Unstable Current Sheets (UCSs). Many well known non linear MHD structures (strong turbulence, current sheet(s), shock(s)) lead asymptotically to the state of turbulent reconnection. We analyze in this article, for the first time, the energization of electrons and ions in a {\bf large scale} environment that {\bf combines} large amplitude disturbances propagating with sub-Alfv\'enic speed with UCSs. The magnetic disturbances interact stochastically (second order Fermi) with the charged particles and they play a crucial role in the heating of the particles, while the UCS interact systematically (first order Fermi) and play a crucial role in the formation of the high energy tail. The synergy of stochastic and systematic acceleration provided by the mixture of magnetic disturbances and UCSs influences the energetics of the thermal and non-thermal particles,
the power law index, and the time the particles remain  inside the energy release volume.
We show that this synergy can explain the observed very fast and impulsive particle acceleration and the slightly delayed formation of a super-hot particle population.
\end{abstract}
\maketitle
\section{Introduction}
Turbulent reconnection can be generated by different, well known, non-linear MHD processes and structures, which
serve as their driver, e.g.\ the evolution of a spectrum of large amplitude MHD waves, the fragmentation of a current sheet(s), or shock(s) (see the reviews of \citealt{Matthaeus11, Cargill12, Lazarian12, Hoshino12, Karimabadi13a, Karimabadi03b, Lazarian15}, and the recent articles of \citealt{Zank15, Matsumoto15} on turbulent reconnection driven by shocks, for a brief, yet incomplete, outline of the relevant literature).

The term ``turbulent reconnection'' appeared first in the article of \cite{Matthaeus86}, and several years later the analytical theory of turbulent reconnection was formulated by \cite{Lazarian99}. In both articles, the role of weak turbulence in the evolution of a reconnecting current sheet was analyzed. In the present article, we expand the term  ``turbulent reconnection'' to denote the co-existence of ``large scale coherent magnetic disturbances'' \citep{Kuramitsu, Greco10, Malapaka13} with ``unstable current sheets'' (UCSs). It has been shown that these two types of nonlinear structures drive and re-enforce each other \citep{Karimabadi03b, Uritsky17}.

There are at least three avenues that lead to the generation of  turbulent reconnection. The first is strong turbulence \citep{Biskamp89, Dmitruk03, Dmitruk04, Arzner06, Isliker17a}. The second is the fragmentation of UCSs: numerous studies have explored the evolution of one or multiple UCSs and have shown that they evolve into a turbulent reconnection environment \citep{Matthaeus86, Lazarian99, Onofri04, Loureiro09,Cassak09, Kowal11, Hoshino12b, Uritsky17}. The third avenue lies downstream of a shock where turbulent reconnection can be driven when the shock is formed in the presence of upstream turbulent flows, e.g.\ as it is the case with the Earth's Bow shock and the solar wind, or with coronal mass ejections traveling inside the turbulent solar wind \citep{Matsumoto15, Zank15, Burgess16}.

Several authors explored the formation of turbulent reconnection in the solar atmosphere, driven by the turbulent convection zone \citep{Parker83, Parker88, Einaudi94, Galsgaard96, Galsgaard97a, Gasgaard97b, Georgoulis98, Rappazzo10, Rappazzo13}. All these studies focused on the formation of UCSs; large amplitude magnetic disturbances were though also present, yet never analyzed in detail until recently \citep{Kontar17}.

The solar wind is probably the most striking example of a turbulently reconnecting plasma flow, and the coexistence of large amplitude magnetic disturbances and UCSs has been analyzed in detail by several authors \citep{Greco10, Osman14, Chasapis15}.

It is natural to expect that in relativistic jets and other astrophysical flows (e.g.\ accretion discs) turbulent reconnection will be
present,
but details have not been worked out yet \citep{Giannios10, Sironi15, Brunetti16}. The acceleration mechanism preferred by most researchers as the best candidate  for explaining the explosive phenomena  in astrophysics remains diffusive shock acceleration, which can be the host of turbulent reconnection \citep{Matsumoto15, Zank15}. The relative importance of the two systematic accelerators (shock acceleration and turbulent reconnection) has not yet been established.

\cite{Karimabadi03b, Uritsky17} pointed out that intermittent plasma turbulence will in general consist of both, coherent structures (UCSs) and large amplitude waves \citep[see also][]{Wang15, Liang16}, and the picture given in \citep{Kowal17} for the MHD evolution of a single small scale UCS. With their simulations they have presented evidence for the excitation of eddies and waves by the motion of fragmented UCSs. They also noted that this complex environment, which we call here turbulent reconnection, heats very efficiently the plasma.

The evolution of an ensemble of charged particles in turbulent reconnection was investigated by several authors using test particle simulations in snapshots of MHD codes \citep{Ambrosiano88, Dmitruk04, Turkmani05, Arzner06, Onofri06, Isliker17a}.

Several authors \citep{Vlahos16, Pisokas16, Isliker17} analyzed the statistical properties of ions and electrons scattered either off large amplitude magnetic disturbances propagating with the Alfv\'en speed (Alfv\' enic Scatterers, ASs), or off UCSs, randomly distributed inside the energy release volume. The interaction of
electrons and ions with the accelerators
is either stochastic when they interact with Alfv\'enic disturbances, as in the model proposed initially by \cite{Fermi49}, or systematic, when they interact with UCSs \citep{Fermi54}. The stochastic energization of ions and electrons leads the initial Maxwellian energy distribution to an asymptotic state and the final distribution is a mixture of a hot and an accelerated plasma, with a power law tail with index $\sim 2$. The acceleration time for parameters related with the solar corona is close to a few seconds \citep[see details in][]{Vlahos16, Pisokas16}.
In the case of systematic acceleration,
when the energy release volume is dominated by UCSs, the particles are mostly accelerated, forming a power law tail with index $\geq 1$ on sub-second time scales, and heating is practically absent \citep[see][]{Vlahos16, Isliker17}.

In most laboratory and astrophysical plasmas the explosive energy release is associated with intense and efficient heating of the bulk of the plasma and with the formation of a power law tail on very fast time scales \citep[see for example the evolution of the photon distribution for solar flares analyzed by][]{Lin03}. \cite{Lin03} pointed out that in the initial rise of a flare substantial particle acceleration is taking place and in the subsequent impulsive phase a coronal super-hot component appears. A possible explanation for the prompt acceleration and the delayed appearance of the super-hot plasma may be related with the differences in the acceleration times between UCSs and the ASs, as reported above. In summary, the synergy of large scale magnetic disturbances and UCSs
in turbulent reconnection
can provide the
explanation
for the appearance of impulsive heating of the super-hot sources and of the non-thermal tails.

In this article, we assume that the energy release volume is in the state of turbulent reconnection and the nonlinear structures (magnetic disturbances and UCSs, interchangeably called here ``active grid points'' or ``scatterers'') are randomly distributed. The charged particles scatter off the active grid points and gain or lose energy. The scatterers are divided into two classes, a fraction P ($0\leq P \leq 1 $) are magnetic disturbances (ASs) and the rest ($1-P$) are UCSs. When $P=0$ all scatterers are UCS and when $P=1$ all scatterers are magnetic disturbances \cite[see][for studies of the extreme cases]{Vlahos16, Pisokas16, Isliker17}.

\section{Mixing stochastic and systematic scatterers}
\subsection{The initial set-up}

The scatterers are randomly chosen and uniformly distributed grid points of a 3D lattice that has linear size $L$ and consists of $(N \times N \times N)$ nodes, with grid size $\ell=L/(N-1)$.
The $N_{\rm sc}$ scatterers form a small fraction $R = N_{\rm sc}/N^3$ of the total number of nodes, and they are
either
ASs
or UCSs, as described above. The mean free path between scatterers can be determined as $\lambda_{\rm sc}=\ell/R$. An ensemble of particles (electrons or ions) are injected into the simulation volume at random grid points, with random direction of motion, and they are allowed to move along the straight lattice edges until they encounter an active grid point. Encounters with scatterers cause a particle to change its direction of motion and to renew its energy by the amount $\Delta W$, which depends on the physical characteristic of the scatterer. This process repeats up to final time or until a particle reaches the lattice boundary and escapes. 
\citep[See][for a more complete description of the model.]{Vlahos16, Pisokas16, Isliker17}

We assume that the simulation volume has length $L = \SI{10000}{km}$, the active grid points ratio is $R = \SIrange{5}{15}{\percent}$, and the injected particles follow a Maxwellian distribution with temperature $T \approx \SI{1e6}{K}$.
If a scatterer is an AS,
the change in energy of a particle amounts to
\begin{equation}\label{e:DW_F2}
    \frac{\Delta W^{\rm (AS)}}{W}\approx\frac{2}{c^2}(V^2-\vec{V} \cdot \vec{u}) ,
\end{equation}
where for head on collisions $\vec{V}\cdot \vec{u} < 0$ and the particle gains energy, for overtaking collisions $\vec{V} \cdot \vec{u} > 0$ and the particle loses energy \citep{Pisokas16}.
The ASs, as stochastic scatterers, transfer energy either to or from an interacting particle, but the overall result for the particles is a gain in energy, with a typical increment of the order of $(\Delta W/W) \approx (V_A/c)^2\sim \num{5e-4}$.

With an UCS as scatterer, the energy gain is caused by the electric field \citep{Kowal11,Isliker17}, and it is given by
\begin{equation} \label{e:DW_EF}
    \Delta W^{\rm (EF)} = |q| E_{\rm eff}\ \ell_{\rm eff}
\end{equation}
where $E_{\rm eff} \approx (V/c) \; \delta B$ is a measure of the effective electric field of the UCS, and $\delta B$ is the fluctuating magnetic field, which is of stochastic nature, as related to the stochastic fluctuations induced by reconnection.
The energy increments in
eq.~\eqref{e:DW_EF} are always positive, as it was shown to hold in different particle in cell simulations \citep{Guo15, Dahlin15, Matsumoto15}, and they do not depend on the instantaneous energy of a particle, instead they are proportional to
the magnetic field fluctuations $\delta B$. The latter
are assumed to
follow a Kolmogorov spectrum, i.e.\ they obey a power law distribution with index \num{5/3} in the range $[\SI{.e-5}{G}, \SI{100}{G}]$. The effective length $\ell_{\rm eff}$ of the interaction of a particle with a UCS is assumed an increasing linear function of $E_{\rm eff}$ (so that small $E_{\rm eff}$ are associated with small scale UCS), restricted to values between \SI{10}{m} and \SI{1}{km} \citep[see more details in][]{Isliker17}.

\subsection{Mixing AS with UCSs dominated by electric fields}

\begin{figure}[ht!]
	\includegraphics[width=0.9\columnwidth]{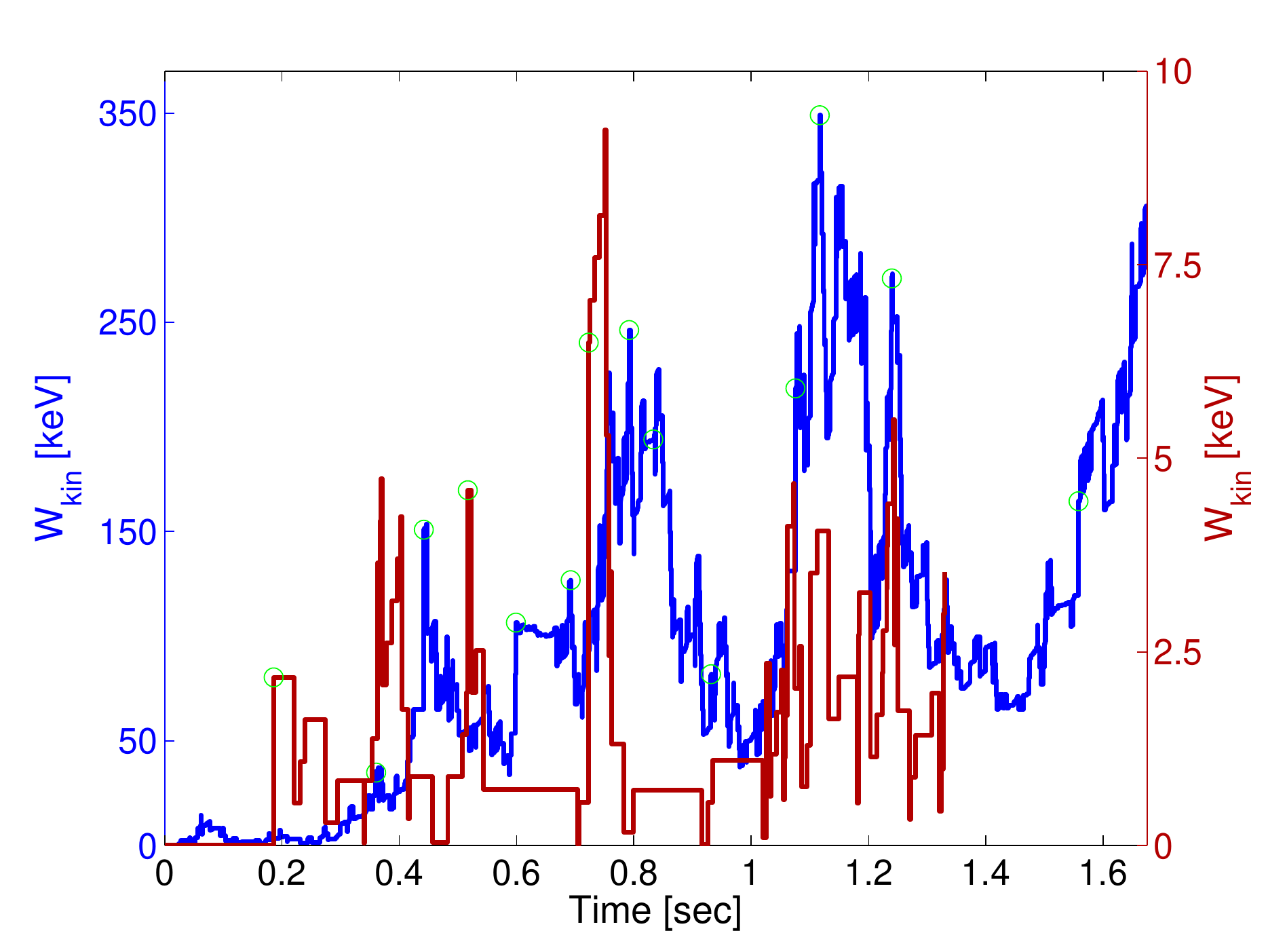}%
	\caption{Kinetic energy of typical electrons as a function of time. 
    } \label{f:Wt}
\end{figure}

In Fig.~\ref{f:Wt} the energy evolution of some typical particles traveling inside a mixture of stochastic and systematic scatterers ($P=0.5$) is shown. The synergy of stochastic acceleration by the AS (classical random walk like behavior) with systematic acceleration at the UCSs (sudden increases of energy) is apparent. 

\begin{figure}[ht]
	\sidesubfloat[]{\includegraphics[width=0.6\columnwidth]{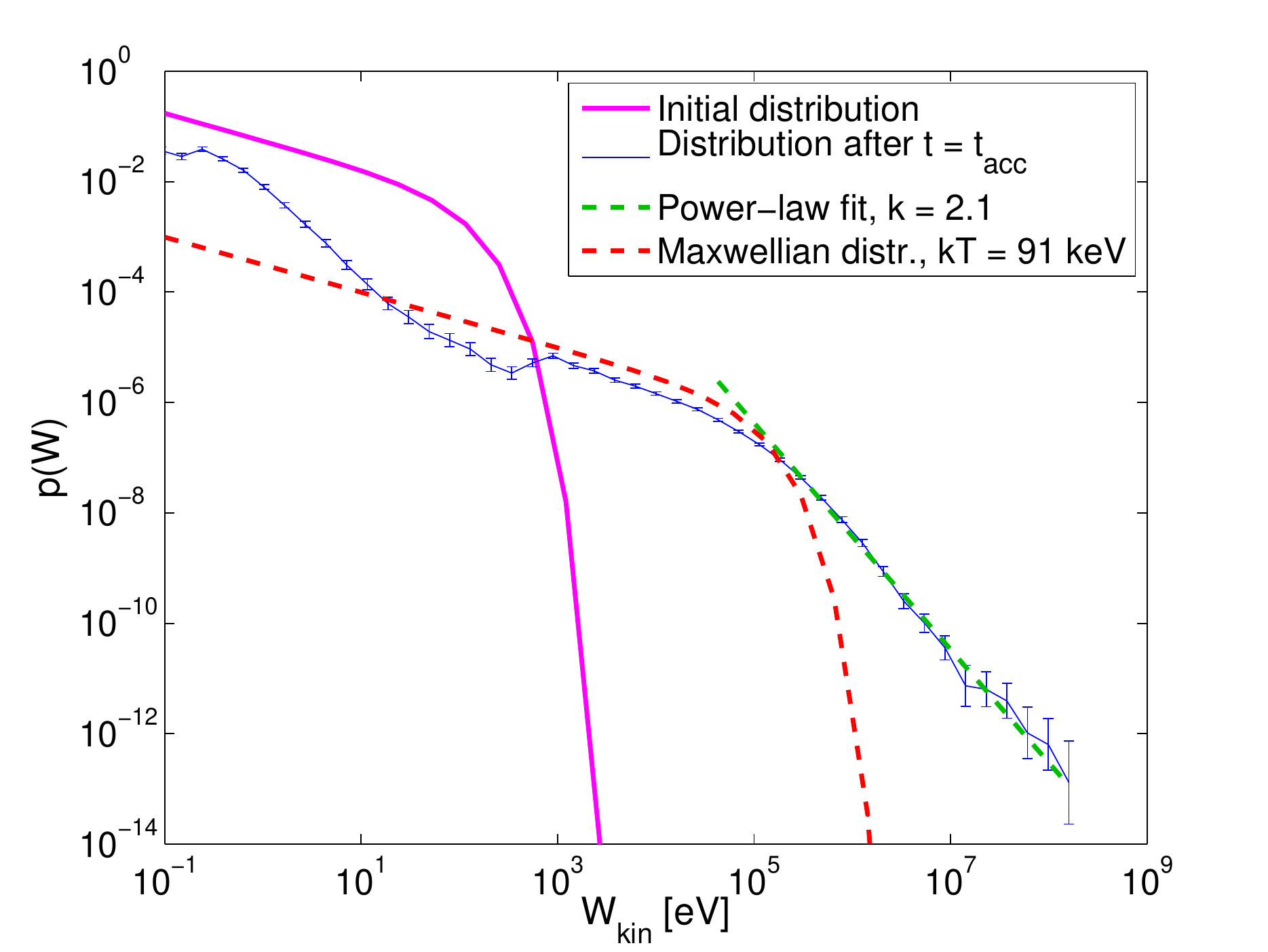}%
		\label{f:AT_tesc:nW50}}\hfill%
	\sidesubfloat[]{\includegraphics[width=0.6\columnwidth]{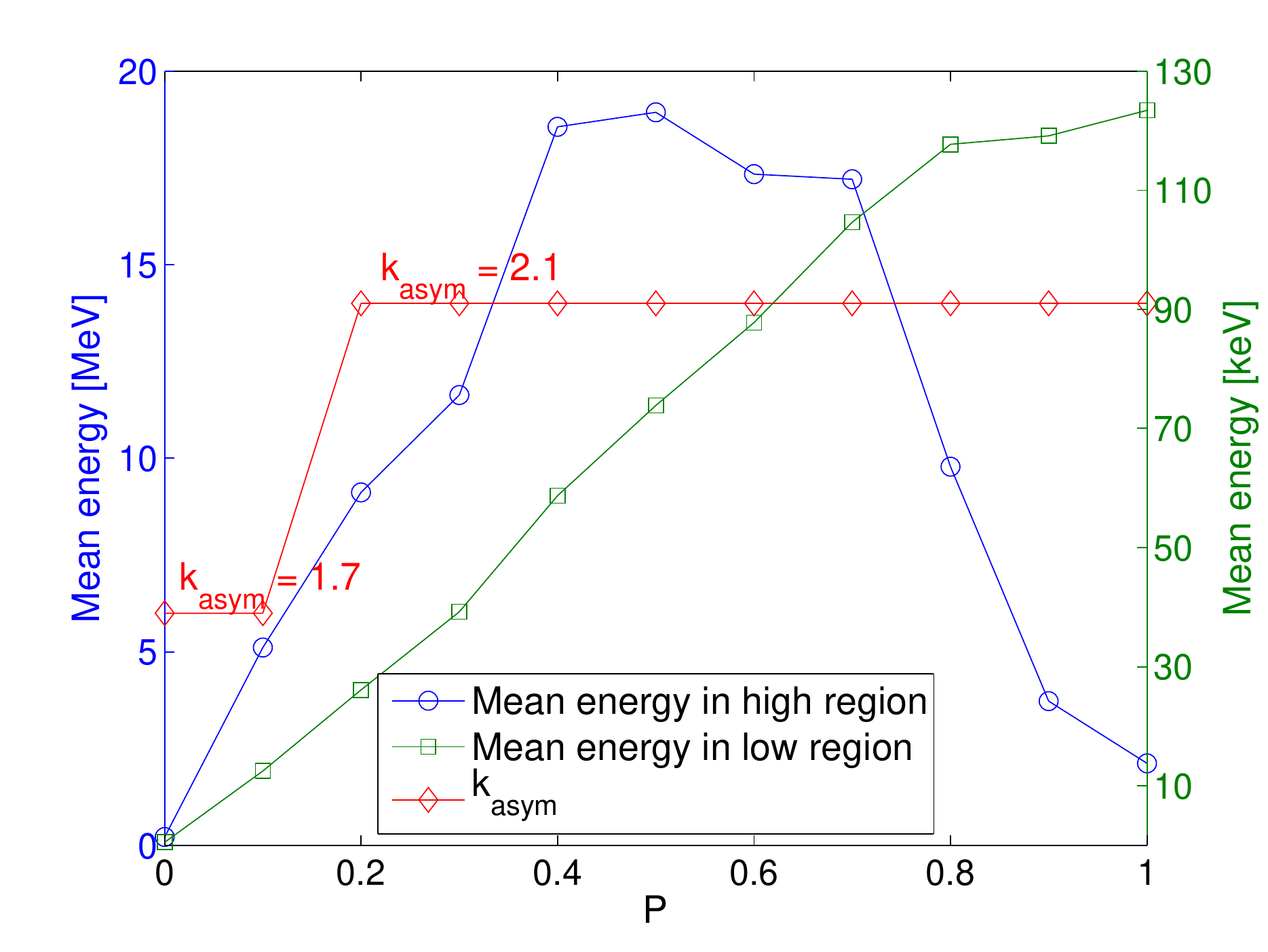}%
		\label{f:AT_tesc:mW_LH}}%
	\caption{
	\protect\subref{f:AT_tesc:nW50} Energy distribution of the electrons that stay inside until $t = t_{\rm acc} \approx \SI{1.7}{sec}$ (blue), for $P = 0.5$; initial distribution (magenta);  Maxwellian fit to the heated low energy region (red dashed), and fit to the power law tail  (green dashed).
	\protect\subref{f:AT_tesc:mW_LH} Mean energy at $t = t_{\rm esc}$ for the high-energy tail (blue) and the heated low-energy region (green) for different ratios $P$ of the two kinds of scatterers;  the red points denote the asymptotic value $k_{\rm asym}$ of the power-law index. 	}
	\label{f:AT_tesc}
\end{figure}

As described above, the particles travel and interact with scatterers until they exit the acceleration volume at some time, which is different for each particle. The median value of these times is the characteristic escape time ($t_{\rm esc}$) for the ensemble, and it coincides with the half time $t_{1/2}$ of the system, defined as the time when half of the initial electron population has escaped.  The energy distribution for the electrons that remain inside the volume exhibits a clear and extended power law tail for $P = 0$, with negligible heating at the low energies, but as $P$ grows, the electrons are also heated under the influence of the magnetic disturbances, until they reach a combination of a hot plasma with a relatively small number of particles in the power law tail for $P = 0.5$, as shown in Fig.~\ref{f:AT_tesc:nW50}. 

The temperature of the heated low energy part of the distribution grows linearly with increasing $P$, starting from a value close to the initial temperature for $P=0$, and reaching a much higher temperature of $\approx \SI{130}{keV}$ for $P=1$ (see Fig.~\ref{f:AT_tesc:mW_LH}). On the other hand, the mean energy of the particles in the high energy tail increases with increasing $P$ until it reaches a maximum value $\approx$ \SIrange{17}{18}{MeV}, forming a plateau for the middle range of $P$ values (\SIrange{0.3}{0.7}{}). For higher $P$ values, the mean energy of the tail drops to $\approx \SI{2}{MeV}$. The synergy of the two classes of scatterers varies the behavior of the system from an efficient accelerator, when the UCSs dominate, to an efficient and excessive heating mechanism combined with acceleration, when both types of scatterers are involved. The power-law tail consists of a small percentage ($\sim$ \SIrange{2}{5}{\percent}) of the total number of particles injected initially
for almost all $P$ values, with the exception of the pure UCS case, $P=0$ \citep{Isliker17}, where the high energy particles are almost \SI{15}{\percent} of the total number of particles at $t = t_{\rm esc}$.

For any combination of the two types of scatterers, the distribution of the high energy particles develops a power law shape, which ultimately attains an asymptotic index $k_{\rm asym}$. The time this occurs is a measure for the acceleration time of the system, $t_{\rm acc}$. According to Fig.~\ref{f:AT_tesc:mW_LH}, when the fraction of the ASs is low ($P < 0.2$), the index $k_{\rm asym}$ is \num{1.7}, as in a pure UCS system, but as the influence of the ASs becomes stronger, we find $k_{\rm asym} \approx 2.1$. The fraction of ASs also affects the acceleration time, which varies around \SIrange{1.5}{2}{sec} for $P \geq 0.2$, but it is much shorter for lower $P$ values, e.g.~$\approx \SI{0.5}{sec}$ for $P = 0.1$ and a few milliseconds for $P = 0$.  Electrons leave the system earlier when ASs are present. We can conclude from this parametric study of the energy distribution that the power law tail is a result of the synergy between the ASs and the UCSs, while the heating of the particles is a sole effect of the ASs.

\subsection{The evolution of the escape time}
\begin{figure}[ht!]
	\sidesubfloat[]{\includegraphics[width=8cm]{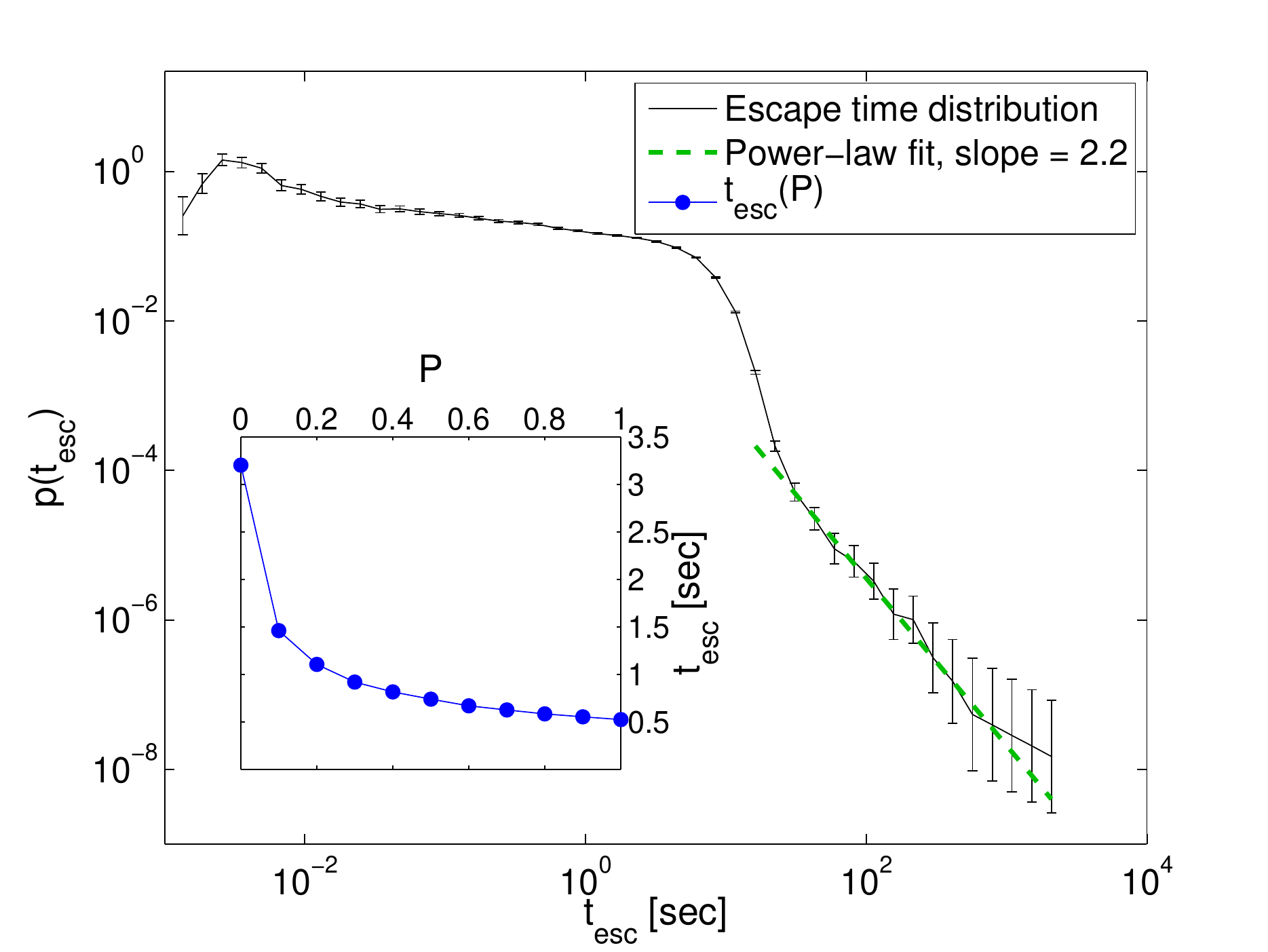}%
		\label{f:esc:nt50}}\\%
	\sidesubfloat[]{\includegraphics[width=8cm]{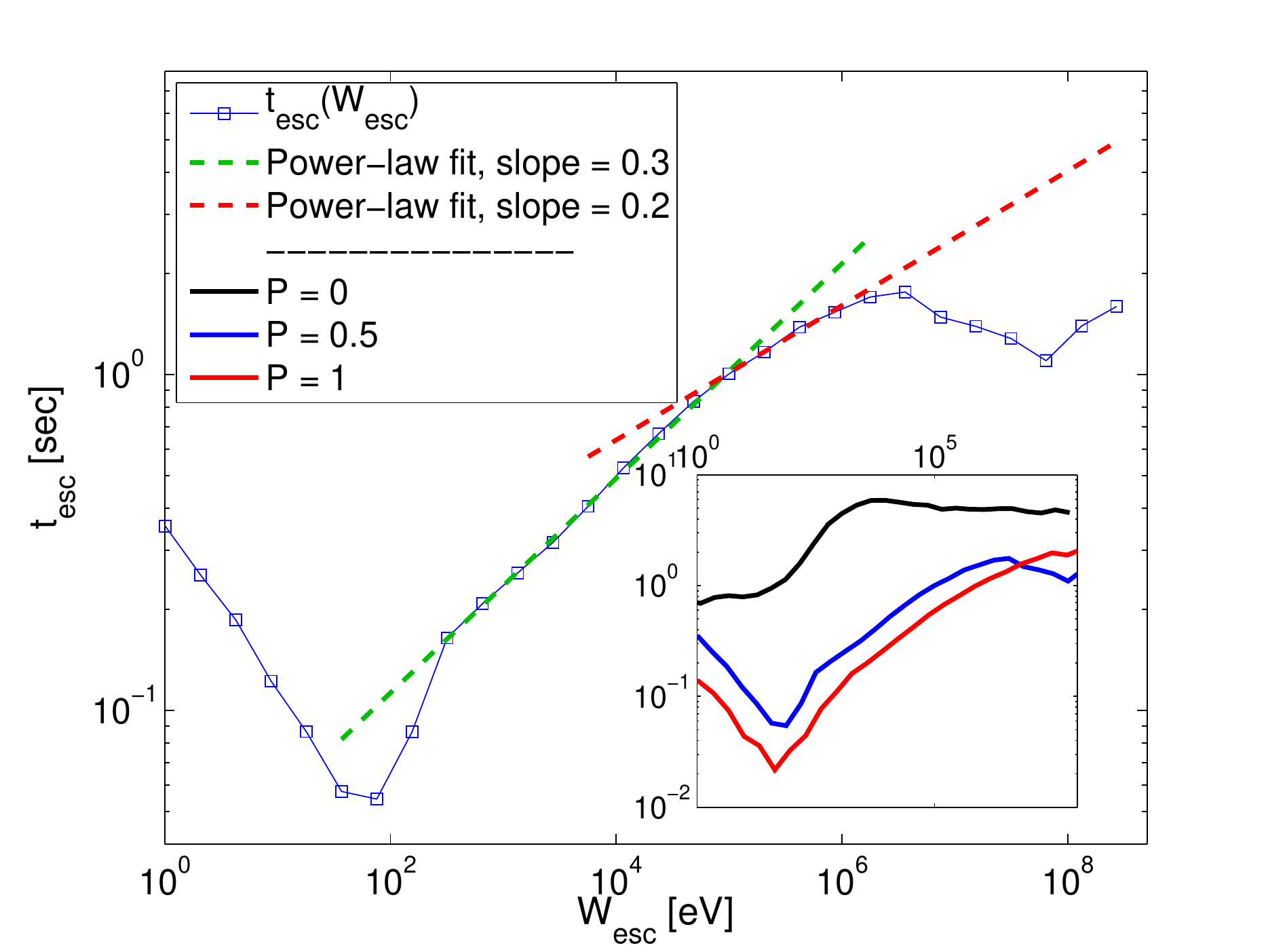}%
		\label{f:esc:tW50}}\\%
	\sidesubfloat[]{\includegraphics[width=8cm]{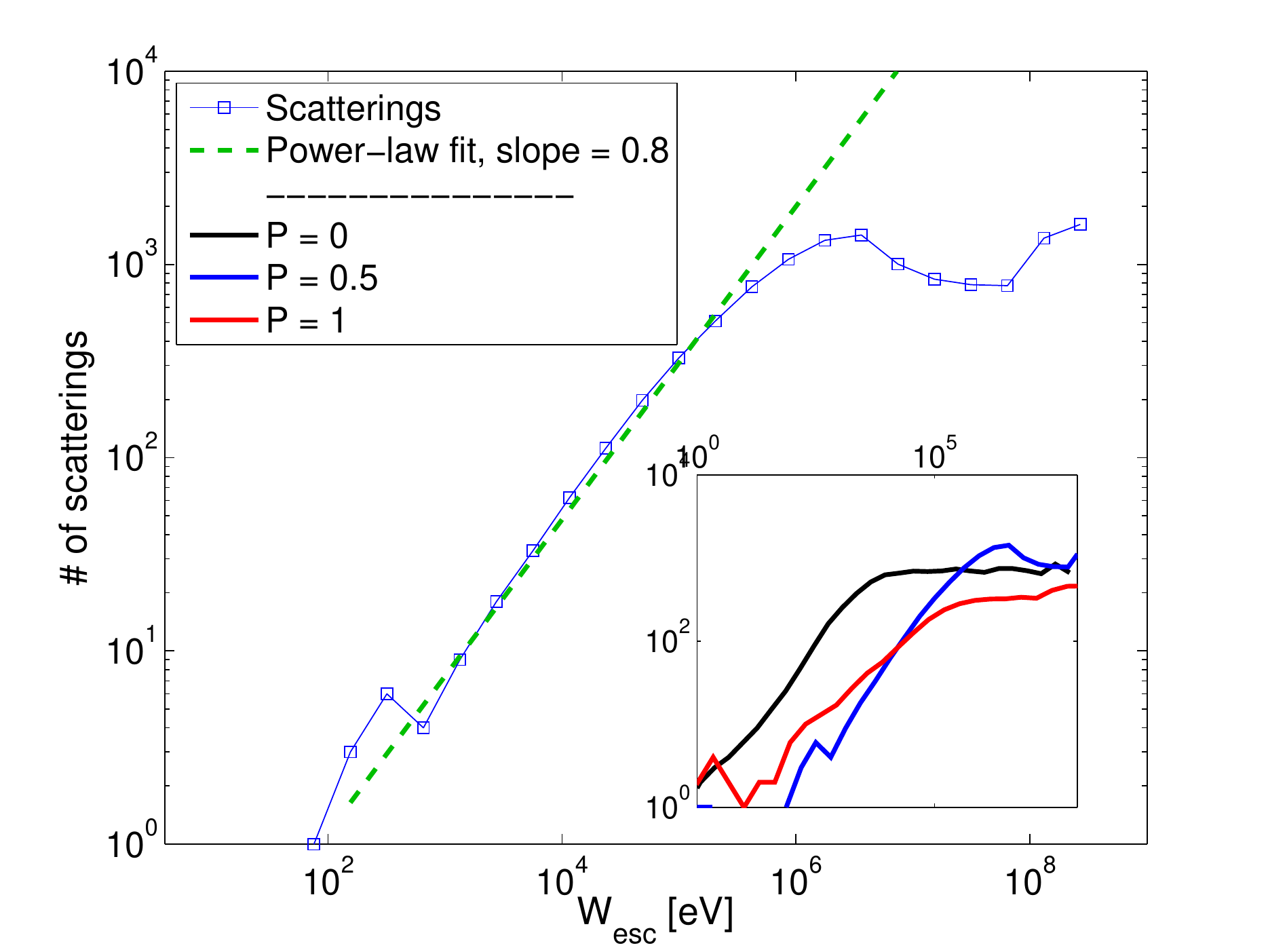}%
		\label{f:esc:kW50}}%
	\caption{\protect\subref{f:esc:nt50} Escape time distribution for $P = 0.5$; \emph{inset:} the escape time distribution for different values of $P$. \protect\subref{f:esc:tW50} The escape time as a function of the escape energy of the electrons for $P = 0.5$; \emph{inset:} comparison between $P = 0$ (black), 0.5 (blue), and 1 (red). \protect\subref{f:esc:kW50} Number of scatterings as a function of the escape energy of the electrons for $P = 0.5$; \emph{inset:} comparison between $P = 0$ (black), 0.5 (blue), and 1 (red).}\label{f:esc}
\end{figure}

As mentioned before, each particle exits the acceleration volume at a different time with a different energy. The escape time distribution
adopts a power law shaped tail, as shown in Fig.~\ref{f:esc:nt50}. 

The distribution of the escape energy, i.e.~the energy with which any particle escapes from the acceleration volume, exhibits the same behavior as the one of the particles that stay inside. The escape time and the escape energy depend on each other; the escape time as a function of the escape energy (determined through binned statistics) is shown in Fig.~\ref{f:esc:tW50} for $P = 0.5$, compared also to the two ``pure'' cases ($P = 0$ and 1). We observe two distinct regions: For  non-relativistic energies, \SIrange{10}{.e5}{eV}, a power law scaling is assumed, $t_{\rm esc} \propto W_{\rm esc}^{0.4}$.

As for the relativistic energies, \SIrange{.e5}{.e9}{eV}, the functional form depends on $P$, e.g. for P=0.5 there is a power law scaling with index 0.2.  In any case, the  escape energy of the electrons is strongly related to the time the particles spend in the system, which is also illustrated when considering the number of scatterings the escaping particles suffer before they escape from the simulation box as a function of their escape energy, see Fig.~\ref{f:esc:kW50}  (determined again through binned statistics), the escape energy increases with increasing number of scatterings, with a saturation at the highest energies.

All of the above results refer to electron populations. When ions are considered, no significant changes are observed. The energy distribution follows the same trends as in the electron case, with corresponding characteristics for the same values of $P$, similarly changing as $P$ varies. Only the time scale is  different, e.g.~the escape time now extends from $\approx \SI{130}{sec}$ for $P = 0$, to $\approx \SI{5}{sec}$ for $P = 0.5$, and it reaches $\approx \SI{3}{sec}$ for $P = 1$. UCSs are not only efficient accelerators, but they also  keep the high energy plasma longer inside the simulation box.

\section{Discussion and Summary}
In most explosive laboratory and astrophysical systems plasma heating is strongly correlated with particle acceleration. In the current literature the problems of heating and particle acceleration are studied separately. In explosive phenomena, particle acceleration has been related with reconnection, shocks, or weak turbulence, but no mechanism for the {\bf impulsive heating} has been proposed so far \citep{Lin03}.

In this article, we have shown that turbulent reconnection can be the solution to the combined problem, since the synergy of ASs and UCSs can provide both, the intense heating and the acceleration of the tail. The mixing of stochastic and systematic acceleration provides an asymptotic energy distribution that is heated substantially and exhibits a power law tail with index $\sim 2$, which is close to the one estimated form the observations. The escape time of the particles depends on the fraction of the two types of scatterers and the energy of the particles. The lower the fraction of Alfv\'enic scatterers,
the larger the escape time of the particles. The most energetic particles stay longest inside the acceleration volume \citep{Krucker08a}.
In the analysis presented in this article, we follow the evolution of the energy distribution for {\bf several seconds} in a {\bf large scale open system}, where the magnetic disturbances and the UCSs have time to bring the energy distribution into an {\bf asymptotic state}.

The scenario proposed here for most explosive phenomena in the solar atmosphere starts with the formation of large scale current sheets, which fragment very quickly, leading to turbulent reconnection. The UCSs  impulsively (in a few milliseconds) accelerate the tail of the energy distribution, and soon after (a few seconds later) the ASs start participating by forming the super-hot sources and reinforcing the tail of the energy distribution \citep[see a clear outline of the observations related to this scenario in][]{Lin03}. The majority of the MHD simulations of explosive phenomena in the solar atmosphere rarely follow the fragmentation of the large scale current sheet, which is usually formed, nor do they consider the generation of magnetic disturbances, due to the limitations in the spatial resolution of these codes. Therefore, the formation of turbulent reconnection and its role in coronal heating and particle acceleration, as analyzed here, has
so-far been ignored.

The spatial transport inside a turbulent plasma will also influence the distribution of the accelerated particles \citep{Bian17}, and the coupling of energy and spatial transport in turbulent reconnecting plasmas remains an open problem. The prompt acceleration and the impulsive heating by turbulent reconnection inside a large scale flaring magnetic topology and the anomalous transport in space will place the simple deterministic scenario for the interpretation of the Hard X-rays during solar flares, as proposed by \citep{Brown71} almost 45 years ago and named ``thick target model'', in a new frame of analysis. We claim that turbulent reconnection will provide answers to many open questions, possibly caused by the simplicity of the physical process adopted in the thick target model.
\begin{acknowledgements}
	This work was supported by the national Programme for the Controlled Thermonuclear Fusion, Hellenic Republic. The sponsors do not bear any responsibility for the content of this work.
\end{acknowledgements}

\begin{thebibliography}{}
\expandafter\ifx\csname natexlab\endcsname\relax\def\natexlab#1{#1}\fi

\bibitem[{Ambrosiano {et~al.}(1988)Ambrosiano, Matthaeus, Goldstein, \&
  Plante}]{Ambrosiano88}
Ambrosiano, J., Matthaeus, W.~H., Goldstein, M.~L., \& Plante, D. 1988, Journal
  of Geophysical Research: Space Physics, 93, 14383

\bibitem[{{Arzner} {et~al.}(2006){Arzner}, {Knaepen}, {Carati}, {Denewet}, \&
  {Vlahos}}]{Arzner06}
{Arzner}, K., {Knaepen}, B., {Carati}, D., {Denewet}, N., \& {Vlahos}, L. 2006,
  Astrophys. J., 637, 322

\bibitem[{{Bian} {et~al.}(2017){Bian}, {Emslie}, \& {Kontar}}]{Bian17}
{Bian}, N.~H., {Emslie}, A.~G., \& {Kontar}, E.~P. 2017, \apj, 835, 262

\bibitem[{{Biskamp} \& {Welter}(1989)}]{Biskamp89}
{Biskamp}, D., \& {Welter}, H. 1989, Physics of Fluids B, 1, 1964

\bibitem[{{Brown}(1971)}]{Brown71}
{Brown}, J.~C. 1971, \solphys, 18, 489

\bibitem[{{Brunetti} \& {Lazarian}(2016)}]{Brunetti16}
{Brunetti}, G., \& {Lazarian}, A. 2016, \mnras, 458, 2584

\bibitem[{{Burgess} {et~al.}(2016){Burgess}, {Gingell}, \&
  {Matteini}}]{Burgess16}
{Burgess}, D., {Gingell}, P.~W., \& {Matteini}, L. 2016, Astrophys. J., 822, 38

\bibitem[{Cargill {et~al.}(2012)Cargill, Vlahos, Baumann, Drake, \&
  Nordlund}]{Cargill12}
Cargill, P., Vlahos, L., Baumann, G., Drake, J., \& Nordlund, {\AA}. 2012,
  Space science reviews, 173, 223

\bibitem[{{Cassak} \& {Drake}(2009)}]{Cassak09}
{Cassak}, P.~A., \& {Drake}, J.~F. 2009, \apjl, 707, L158

\bibitem[{{Chasapis} {et~al.}(2015){Chasapis}, {Retin{\`o}}, {Sahraoui},
  {Vaivads}, {Khotyaintsev}, {Sundkvist}, {Greco}, {Sorriso-Valvo}, \&
  {Canu}}]{Chasapis15}
{Chasapis}, A., {Retin{\`o}}, A., {Sahraoui}, F., {et~al.} 2015, \apjl, 804, L1

\bibitem[{{Dahlin} {et~al.}(2015){Dahlin}, {Drake}, \& {Swisdak}}]{Dahlin15}
{Dahlin}, J.~T., {Drake}, J.~F., \& {Swisdak}, M. 2015, Physics of Plasmas, 22,
  100704

\bibitem[{Dmitruk {et~al.}(2003)Dmitruk, Matthaeus, Seenu, \&
  Brown}]{Dmitruk03}
Dmitruk, P., Matthaeus, W., Seenu, N., \& Brown, M.~R. 2003, The Astrophysical
  Journal Letters, 597, L81

\bibitem[{{Dmitruk} {et~al.}(2004){Dmitruk}, {Matthaeus}, \&
  {Seenu}}]{Dmitruk04}
{Dmitruk}, P., {Matthaeus}, W.~H., \& {Seenu}, N. 2004, Astrophys. J., 617, 667

\bibitem[{{Einaudi} \& {Velli}(1994)}]{Einaudi94}
{Einaudi}, G., \& {Velli}, M. 1994, Space Sci. Rev., 68, 97

\bibitem[{{Fermi}(1949)}]{Fermi49}
{Fermi}, E. 1949, Physical Review, 75, 1169

\bibitem[{{Fermi}(1954)}]{Fermi54}
---. 1954, \apj, 119, 1

\bibitem[{{Galsgaard} \& {Nordlund}(1996)}]{Galsgaard96}
{Galsgaard}, K., \& {Nordlund}, {\AA}. 1996, Journal Geoph. Res, 101, 13445

\bibitem[{{Galsgaard} \& {Nordlund}(1997{\natexlab{a}})}]{Galsgaard97a}
---. 1997{\natexlab{a}}, Journal Geoph. Res., 102, 219

\bibitem[{{Galsgaard} \& {Nordlund}(1997{\natexlab{b}})}]{Gasgaard97b}
---. 1997{\natexlab{b}}, Journal Geoph. Res., 102, 231

\bibitem[{{Georgoulis} {et~al.}(1998){Georgoulis}, {Velli}, \&
  {Einaudi}}]{Georgoulis98}
{Georgoulis}, M.~K., {Velli}, M., \& {Einaudi}, G. 1998, Astrophys. J., 497,
  957

\bibitem[{{Giannios}(2010)}]{Giannios10}
{Giannios}, D. 2010, \mnras, 408, L46

\bibitem[{{Greco} {et~al.}(2010){Greco}, {Perri}, \& {Zimbardo}}]{Greco10}
{Greco}, A., {Perri}, S., \& {Zimbardo}, G. 2010, Journal of Geophysical
  Research (Space Physics), 115, A02203

\bibitem[{{Guo} {et~al.}(2015){Guo}, {Liu}, {Daughton}, \& {Li}}]{Guo15}
{Guo}, F., {Liu}, Y.-H., {Daughton}, W., \& {Li}, H. 2015, \apj, 806, 167

\bibitem[{{Hoshino}(2012)}]{Hoshino12b}
{Hoshino}, M. 2012, Physical Review Letters, 108, 135003

\bibitem[{{Hoshino} \& {Lyubarsky}(2012)}]{Hoshino12}
{Hoshino}, M., \& {Lyubarsky}, Y. 2012, \ssr, 173, 521

\bibitem[{{Isliker} {et~al.}(2017{\natexlab{a}}){Isliker}, {Pisokas}, {Vlahos},
  \& {Anastasiadis}}]{Isliker17}
{Isliker}, H., {Pisokas}, T., {Vlahos}, L., \& {Anastasiadis}, A.
  2017{\natexlab{a}}, ArXiv e-prints, arXiv:1709.08269

\bibitem[{{Isliker} {et~al.}(2017{\natexlab{b}}){Isliker}, {Vlahos}, \&
  {Constantinescu}}]{Isliker17a}
{Isliker}, H., {Vlahos}, L., \& {Constantinescu}, D. 2017{\natexlab{b}},
  Physical Review Letters, 119, 045101

\bibitem[{Karimabadi {et~al.}(2013)Karimabadi, Roytershteyn, Daughton, \&
  Liu}]{Karimabadi13a}
Karimabadi, H., Roytershteyn, V., Daughton, W., \& Liu, Y.-H. 2013, Space
  Science Reviews, 178, 307

\bibitem[{{Karimabadi} {et~al.}(2013){Karimabadi}, {Roytershteyn}, {Wan},
  {Matthaeus}, {Daughton}, {Wu}, {Shay}, {Loring}, {Borovsky}, {Leonardis},
  {Chapman}, \& {Nakamura}}]{Karimabadi03b}
{Karimabadi}, H., {Roytershteyn}, V., {Wan}, M., {et~al.} 2013, Physics of
  Plasmas, 20, 012303

\bibitem[{{Kontar} {et~al.}(2017){Kontar}, {Perez}, {Harra}, {Kuznetsov},
  {Emslie}, {Jeffrey}, {Bian}, \& {Dennis}}]{Kontar17}
{Kontar}, E.~P., {Perez}, J.~E., {Harra}, L.~K., {et~al.} 2017, Physical Review
  Letters, 118, 155101

\bibitem[{{Kowal} {et~al.}(2011){Kowal}, {de Gouveia Dal Pino}, \&
  {Lazarian}}]{Kowal11}
{Kowal}, G., {de Gouveia Dal Pino}, E.~M., \& {Lazarian}, A. 2011, \apj, 735,
  102

\bibitem[{{Kowal} {et~al.}(2017){Kowal}, {Falceta-Gon{\c c}alves}, {Lazarian},
  \& {Vishniac}}]{Kowal17}
{Kowal}, G., {Falceta-Gon{\c c}alves}, D.~A., {Lazarian}, A., \& {Vishniac},
  E.~T. 2017, \apj, 838, 91

\bibitem[{{Krucker} \& {Lin}(2008)}]{Krucker08a}
{Krucker}, S., \& {Lin}, R.~P. 2008, \apj, 673, 1181

\bibitem[{{Kuramitsu} \& {Hada}(2000)}]{Kuramitsu}
{Kuramitsu}, Y., \& {Hada}, T. 2000, \grl, 27, 629

\bibitem[{{Lazarian} {et~al.}(2015){Lazarian}, {Eyink}, {Vishniac}, \&
  {Kowal}}]{Lazarian15}
{Lazarian}, A., {Eyink}, G., {Vishniac}, E., \& {Kowal}, G. 2015, Philosophical
  Transactions of the Royal Society of London Series A, 373, 20140144

\bibitem[{{Lazarian} \& {Vishniac}(1999)}]{Lazarian99}
{Lazarian}, A., \& {Vishniac}, E.~T. 1999, Astrophys. J., 517, 700

\bibitem[{Lazarian {et~al.}(2012)Lazarian, Vlahos, Kowal, Yan, Beresnyak, \&
  Dal~Pino}]{Lazarian12}
Lazarian, A., Vlahos, L., Kowal, G., {et~al.} 2012, Space science reviews, 173,
  557

\bibitem[{{Liang} {et~al.}(2016){Liang}, {Lin}, {Johnson}, {Wang}, \&
  {Wang}}]{Liang16}
{Liang}, J., {Lin}, Y., {Johnson}, J.~R., {Wang}, X., \& {Wang}, Z.-X. 2016,
  Journal of Geophysical Research (Space Physics), 121, 6526

\bibitem[{{Lin} {et~al.}(2003){Lin}, {Krucker}, {Hurford}, {Smith}, {Hudson},
  {Holman}, {Schwartz}, {Dennis}, {Share}, {Murphy}, {Emslie}, {Johns-Krull},
  \& {Vilmer}}]{Lin03}
{Lin}, R.~P., {Krucker}, S., {Hurford}, G.~J., {et~al.} 2003, \apjl, 595, L69

\bibitem[{{Loureiro} {et~al.}(2009){Loureiro}, {Uzdensky}, {Schekochihin},
  {Cowley}, \& {Yousef}}]{Loureiro09}
{Loureiro}, N.~F., {Uzdensky}, D.~A., {Schekochihin}, A.~A., {Cowley}, S.~C.,
  \& {Yousef}, T.~A. 2009, \mnras, 399, L146

\bibitem[{{Malapaka} \& {M{\"u}ller}(2013)}]{Malapaka13}
{Malapaka}, S.~K., \& {M{\"u}ller}, W.-C. 2013, \apj, 778, 21

\bibitem[{{Matsumoto} {et~al.}(2015){Matsumoto}, {Amano}, {Kato}, \&
  {Hoshino}}]{Matsumoto15}
{Matsumoto}, Y., {Amano}, T., {Kato}, T.~N., \& {Hoshino}, M. 2015, Science,
  347, 974

\bibitem[{{Matthaeus} \& {Lamkin}(1986)}]{Matthaeus86}
{Matthaeus}, W.~H., \& {Lamkin}, S.~L. 1986, Physics of Fluids, 29, 2513

\bibitem[{{Matthaeus} \& {Velli}(2011)}]{Matthaeus11}
{Matthaeus}, W.~H., \& {Velli}, M. 2011, Space Science Reviews, 160, 145

\bibitem[{{Onofri} {et~al.}(2006){Onofri}, {Isliker}, \& {Vlahos}}]{Onofri06}
{Onofri}, M., {Isliker}, H., \& {Vlahos}, L. 2006, Physical Review Letters, 96,
  151102

\bibitem[{{Onofri} {et~al.}(2004){Onofri}, {Primavera}, {Malara}, \&
  {Veltri}}]{Onofri04}
{Onofri}, M., {Primavera}, L., {Malara}, F., \& {Veltri}, P. 2004, Physics of
  Plasmas, 11, 4837

\bibitem[{{Osman} {et~al.}(2014){Osman}, {Matthaeus}, {Gosling}, {Greco},
  {Servidio}, {Hnat}, {Chapman}, \& {Phan}}]{Osman14}
{Osman}, K.~T., {Matthaeus}, W.~H., {Gosling}, J.~T., {et~al.} 2014, Physical
  Review Letters, 112, 215002

\bibitem[{{Parker}(1983)}]{Parker83}
{Parker}, E.~N. 1983, Astrophys. J., 264, 635

\bibitem[{{Parker}(1988)}]{Parker88}
---. 1988, Astrophys. J., 330, 474

\bibitem[{{Pisokas} {et~al.}(2017){Pisokas}, {Vlahos}, {Isliker}, {Tsiolis}, \&
  {Anastasiadis}}]{Pisokas16}
{Pisokas}, T., {Vlahos}, L., {Isliker}, H., {Tsiolis}, V., \& {Anastasiadis},
  A. 2017, \apj, 835, 214

\bibitem[{{Rappazzo} {et~al.}(2010){Rappazzo}, {Velli}, \&
  {Einaudi}}]{Rappazzo10}
{Rappazzo}, A.~F., {Velli}, M., \& {Einaudi}, G. 2010, Astroph. J., 722, 65

\bibitem[{{Rappazzo} {et~al.}(2013){Rappazzo}, {Velli}, \&
  {Einaudi}}]{Rappazzo13}
---. 2013, Astroph. J., 771, 76

\bibitem[{{Sironi} {et~al.}(2015){Sironi}, {Petropoulou}, \&
  {Giannios}}]{Sironi15}
{Sironi}, L., {Petropoulou}, M., \& {Giannios}, D. 2015, \mnras, 450, 183

\bibitem[{Turkmani {et~al.}(2005)Turkmani, Vlahos, Galsgaard, Cargill, \&
  Isliker}]{Turkmani05}
Turkmani, R., Vlahos, L., Galsgaard, K., Cargill, P., \& Isliker, H. 2005, The
  Astrophysical Journal Letters, 620, L59

\bibitem[{{Uritsky} {et~al.}(2017){Uritsky}, {Roberts}, {DeVore}, \&
  {Karpen}}]{Uritsky17}
{Uritsky}, V.~M., {Roberts}, M.~A., {DeVore}, C.~R., \& {Karpen}, J.~T. 2017,
  \apj, 837, 123

\bibitem[{{Vlahos} {et~al.}(2016){Vlahos}, {Pisokas}, {Isliker}, {Tsiolis}, \&
  {Anastasiadis}}]{Vlahos16}
{Vlahos}, L., {Pisokas}, T., {Isliker}, H., {Tsiolis}, V., \& {Anastasiadis},
  A. 2016, Astrophys. J, 827, L3

\bibitem[{{Wang} {et~al.}(2015){Wang}, {Li}, {Ma}, {Zhang}, \& {Lee}}]{Wang15}
{Wang}, L.~C., {Li}, L.~J., {Ma}, Z.~W., {Zhang}, X., \& {Lee}, L.~C. 2015,
  Physics Letters A, 379, 2068

\bibitem[{{Zank} {et~al.}(2015){Zank}, {Hunana}, {Mostafavi}, {Le Roux}, {Li},
  {Webb}, {Khabarova}, {Cummings}, {Stone}, \& {Decker}}]{Zank15}
{Zank}, G.~P., {Hunana}, P., {Mostafavi}, P., {et~al.} 2015, \apj, 814, 137

\end{thebibliography}

\end{document}